\documentclass[aps,superscriptaddress,reprint]{revtex4-1}
\usepackage{amsfonts}
\usepackage{amsmath}
\usepackage{amssymb}
\usepackage{graphicx}
\usepackage{bm}
\usepackage{xcolor}

\renewcommand{\d}{\text{d}}

\begin{document}

\title{Hydrodynamic response of a surfactant-laden interface to a radial flow}
\author{T. Bickel}
\email{thomas.bickel@u-bordeaux.fr}
\affiliation{Univ. Bordeaux, CNRS, Laboratoire Ondes et Mati\`ere d'Aquitaine (UMR 5798), F-33400 Talence, France}
\author{J.-C. Loudet}
\affiliation{Univ. Bordeaux, CNRS, Centre de Recherche Paul Pascal (UMR 5031), F-33600 Pessac, France}
\affiliation{Department of Mathematics, University of British Columbia, Vancouver, BC V6T 1Z3, Canada}
\author{G. Koleski}
\affiliation{Univ. Bordeaux, CNRS, Laboratoire Ondes et Mati\`ere d'Aquitaine (UMR 5798), F-33400 Talence, France}
\affiliation{Univ. Bordeaux, CNRS, Centre de Recherche Paul Pascal (UMR 5031), F-33600 Pessac, France}
\author{B. Pouligny}
\affiliation{Univ. Bordeaux, CNRS, Centre de Recherche Paul Pascal (UMR 5031), F-33600 Pessac, France}

\begin{abstract}
We study the features of a radial Stokes flow due to a submerged jet directed toward a liquid-air interface. 
The presence of surface-active impurities confers to the interface an in-plane elasticity that resists the incident flow. 
Both analytical and numerical calculations show that a minute amount of surfactants is enough to profoundly alter  the  morphology of the flow. The hydrodynamic response of the interface is affected as well, shifting from slip to no-slip boundary condition as the surface compressibility decreases. We argue that the competition between the divergent outward flow and the elastic response of the interface may actually be used as a practical way to detect and quantify a small amount of impurities. 
\end{abstract}

\maketitle

\section{Introduction}

Contamination of the water-air interface is a long-standing issue of interfacial science~\cite{kimNatPhys2017,arangalangeSM2018,uematsuCOE2019}. Because of its high surface  tension, an aqueous interface is susceptible to adsorption of surface-active impurities that are inevitably present in the environment. While traces of surfactants are generally difficult to detect by conventional methods,  interfacial stresses due to a minute amount of surfactants have the capacity to strongly affect the hydrodynamic response of a liquid.
It was for instance recognized that the retarded motion of a  bubble rising in a liquid is due to the presence of impurities~\cite{levichbook,takagiARFM2011}.
Likewise, it has long been known that a small amount of surfactants has a stabilizing effect on convective instabilities~\cite{bergCES1965}. Surface contamination is also suspected to affect the morphology of  ``coffee ring'' patterns observed in droplet evaporation experiments~\cite{deeganNature1997,huJPCB2006, kimPRL2016}.

As the size of the system decreases, interfacial contributions become increasingly relevant.
Microfluidic experiments revealed for instance that traces of surfactants can severely limit the drag reduction of superhydrophobic surfaces~\cite{peaudecerfPNAS2017}.
Impurities at the water-air interface have also been shown to affect its viscoelastic response in AFM experiments~\cite{manorPRL2008,maaliPRL2017}.
Other experiments suggest that surface-active contaminants can promote the rupture of $\mu$m-thick free-standing  films~\cite{neelJFM2018}. At even smaller scales, the stability of surface nanobubbles is  attributed to the presence of impurities~\cite{duckerLangmuir2009,dasPRE2010}, whereas nanomolar concentrations of charged contaminants are invoked in order to explain anomalous surface tension variations  (Jones-Ray effect) in electrolyte solutions~\cite{uematsuJPCL2018}.

These selected examples illustrate the ubiquity of contaminants and the need to take them into consideration when dealing with free surface flows. As a matter of fact, the chemical nature of impurities and their concentrations are likely to vary from experiment to experiment. Indeed, water can be polluted during the preparation or  during the experiment itself, and, given the various cleaning procedures, the nature of the contaminants is largely unknown. Still, the water-air interface remains a popular experimental model system. There is therefore a need  to quantify the presence of surface-active agents at extremely low concentration.

In the present work, we report on the features of the flow due to a submerged jet directed toward the interface of a viscous liquid. If the system is perfectly clean,  the interface is stress-free and the hydrodynamic boundary condition corresponds to perfect slip. The situation gets more involved when a dilute monolayer of surface-active species is irreversibly adsorbed at the interface. Indeed, the convective sweeping of the surfactants by the radial flow forces them to accumulate at the boundaries of the experimental cell. The ensuing tension gradient then gives rise to restoring Marangoni forces which oppose the centrifugal flow. This mechanism thus provides an elastic feature to the interface. Eventually, the surface becomes so rigid that the inward and outward flows cancel exactly, resulting in an effective no-slip boundary condition at the interface. 

From a mathematical viewpoint, the hydrodynamic response of the interface can be expressed as a mixed boundary value problem, which is known as the stagnant cap model in the context of translating bubbles~\cite{levichbook,palaJFM2006}. This  formulation eventually reduces the transport equations to a set of dual integral equations. The latter, which are  discussed in several textbooks~\cite{sneddonbook,duffybook},  are also commonly found in the fluid mechanics literature. Recent applications include for instance the motion of a disk through a rotating fluid~\cite{tanzoshJFM1995}, the Marangoni propulsion of a thin disk at a liquid interface~\cite{laugaJFM2012}, or the self-phoretic actuation of Janus particles~\cite{bickelPRE2013}.
Here, we construct an exact solution    by first 
using Hankel transforms in order to eliminate radial derivatives. Provided that surface diffusion can be neglected,  the mixed boundary value problem then leads to a set of  integral equations with Bessel function kernels.

The remaining of the paper is organized as follows. We first describe in Sec.~\ref{model} the theoretical model, which, given some legitimate approximations, is analytically solved in Sec.~\ref{analytical}. Details regarding the calculations, in particular concerning  dual integral equations, are discussed in Appendix~\ref{appAq}. We then compare the analytical predictions with the results of numerical simulations in Sec.~\ref{numerical}. The outcomes of this work are finally summarized and discussed in Sec.~\ref{discussion}.

\begin{figure}
\includegraphics[width=0.9\columnwidth]{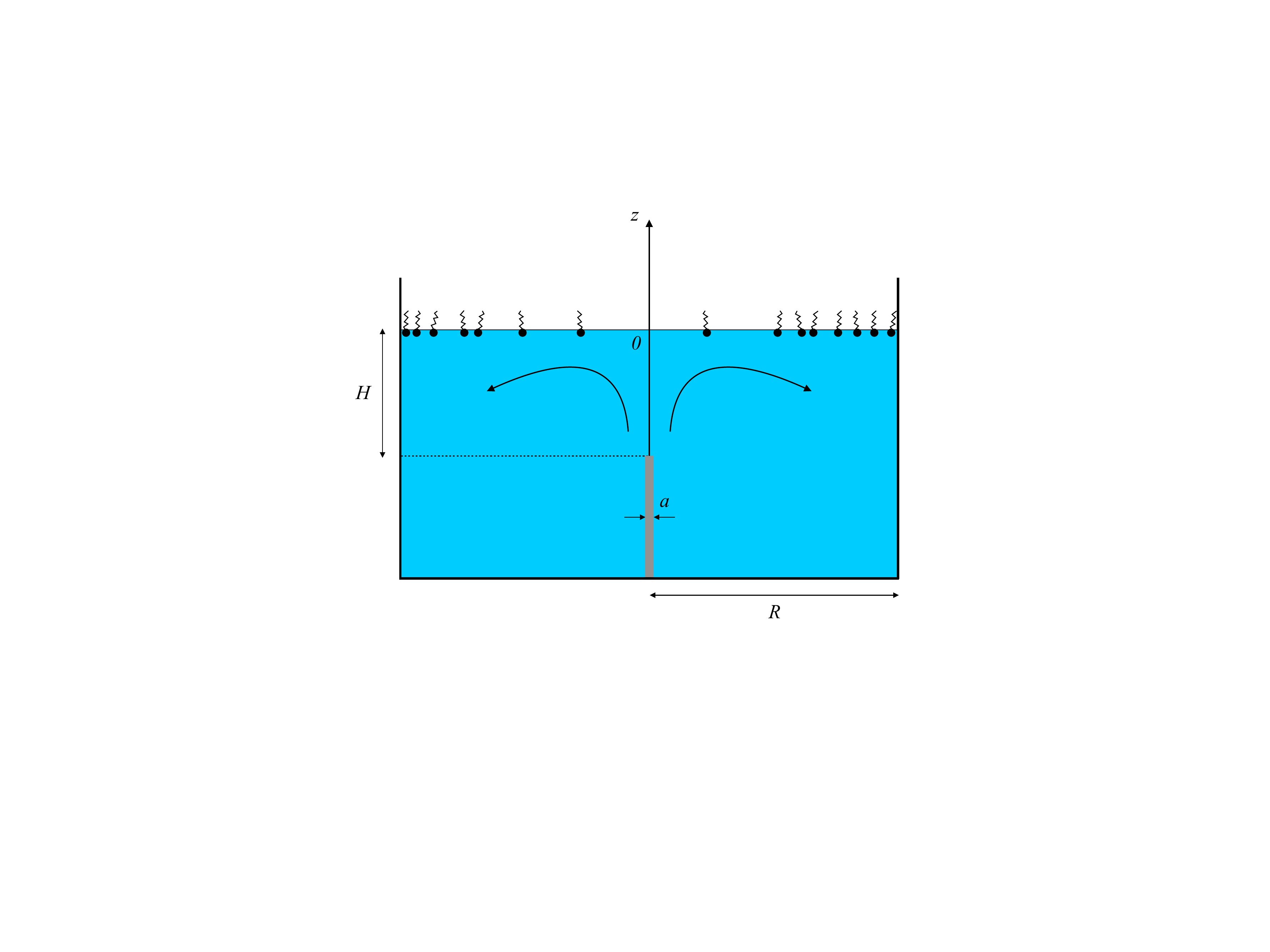}
\caption{Schematic representation of the system. A submerged jet is directed toward an interface covered with insoluble surface-active impurities (represented as surfactants). Impurities are then swept away toward the cell boundaries, which induces a Marangoni counterflow. The system is  invariant by rotation around the $z$-axis.}
\label{fig1}
\end{figure}

\section{General considerations}
\label{model}

The situation under investigation is schematically drawn in Fig.~\ref{fig1}. A newtonian, incompressible liquid of viscosity~$\eta$ and mass density~$\rho$ is enclosed in a cylindrical cell of radius~$R$. A submerged jet of the same liquid is injected through a narrow  tube of opening radius~$a$, whose extremity lies at  distance~$H$ below the interface. The axis of the tube is vertical and coincides with the axis of the cylinder.
The free  interface is horizontal and located at $z=0$, the liquid phase extending in the region $z<0$.
 The system is rotationally invariant so that we set  $\mathbf{r}=(r,z)$, with $r=\sqrt{x^2+y^2}$.

For the sake of simplicity, we neglect fluid inertia and focus on the Stokes regime of the flow. The velocity and pressure fields are then solution of the incompressible Stokes equations
\begin{equation}
\eta \nabla^2 \mathbf{v} = \bm {\nabla} p \ , \quad \text{and} \quad
\bm{\nabla} \cdot \mathbf{v} =0 \ . \label{stokes}
\end{equation}
The discussion is also restricted to the regime of asymptotically small capillary number $\text{Ca} = \eta V_0 / \gamma \ll 1$, with  $\gamma$ the surface tension  and $V_0$ a characteristic speed to be specified below. This condition, which is readily satisfied for velocities pertaining to the viscous regime, implies that the free interface is not deformed by the jet. 
The normal component of the velocity then vanishes at the interface
\begin{equation}
v_z \big\vert_{z=0} = 0 \ .
\label{bcint}
\end{equation}

Our main goal is to elucidate the features of the flow when surface-active molecules are irreversibly adsorbed  at the free interface.
In response to liquid injection, the molecules are swept toward the periphery of the cell, where they accumulate and lower the surface tension.
The surface concentration $\Gamma (r,t)$ of insoluble surfactants then obeys the advection-diffusion equation
\begin{equation}
\frac{\partial \Gamma}{\partial t} + \frac{1}{r} \frac{\partial}{\partial r}  \Big( r v_r(r,0) \Gamma \Big)=  \frac{D}{r} \frac{\partial}{\partial r} \left( r \frac{\partial \Gamma}{\partial r}  \right) \ ,
\label{adeq}
\end{equation}
with $D$ the  diffusion coefficient along the interface. The equilibrium concentration (\textit{i.e.}, in the absence of flow) is denoted~$\Gamma_0$.
In this equation, the relative contribution of diffusion and advection  is quantified by the  P\'eclet number $\text{Pe} = H V_0/D$.
The diffusion coefficient ranges from $D \sim 10^{-9}$~m$^{2}\cdot$s$^{-1}$ for smaller surfactant molecules, up  to $D \sim 10^{-12}$~m$^{2}\cdot$s$^{-1}$ for larger contaminants.  The  P\'eclet number is thus expected to remain very high for length or velocity scales up to the millimeter range, for which $\text{Pe} >10^3$.
As a consequence, the transport of surfactant molecules is primarily controlled by advection.

In general, the surface tension~$\gamma$ is a decreasing function of the local surfactant concentration.
The interfacial velocity and concentration fields are then coupled through the Marangoni boundary condition
\begin{equation}
\eta \frac{ \partial v_r}{\partial z} \bigg\vert_{z=0} = \frac{\partial \gamma}{\partial r}   \ .
\label{bcMar}
\end{equation}
This relation states that an inhomogeneity of surface tension  induces a shear stress at the interface, therefore leading to a flow in the aqueous phase~\cite{scrivenNature1960}.
To discuss interfacial stresses, it is convenient to introduce the surface pressure $\Pi(\Gamma)=\gamma_0 - \gamma(\Gamma)$, with $\gamma_0=\gamma(\Gamma=0)$ the surface tension of the clean interface. A key ingredient of the analysis is then provided by the equation of state that relates~$\Pi$ and~$\Gamma$, or, equivalently, by the Gibbs elasticity coefficient defined as $E =  \Gamma \left(  \partial \Pi/\partial \Gamma\right)$~\cite{langevinARFM2014}.
Here we adopt  the Langmuir adsorption model~\cite{kralchevsky2009}
\begin{equation}
E=\frac{\Gamma k_B T}{1-\Gamma/\Gamma_\infty} \ ,
\label{elaslangmuir}
\end{equation}
with $k_B$ the Boltzmann constant, $T$ the absolute temperature,  and  $\Gamma_\infty$ the concentration at saturation. The latter accounts for the finite area occupied by individual surfactant molecules. Typical values for the maximum packing concentration are of the order of $\Gamma_\infty \sim 10^{6}$~molecules$\cdot \mu$m$^{-2}$~\cite{kralchevsky2009}.
In the dilute limit $\Gamma \ll\Gamma_\infty$, the Gibbs elasticity grows linearly with the concentration, $E=\Gamma k_BT$.
Nonlinear contributions then become increasingly relevant, and  the Gibbs elasticity eventually diverges in the incompressible limit $\Gamma \to \Gamma_{\infty}$.

Accumulation of surfactants at the periphery of the domain  results in the stiffening of the  interface.
The issue is then to quantify the competition between the applied shear stress and the resisting surface elasticity.
Since emphasis is put on the dilute regime, the reference Gibbs elasticity is set by the equilibrium value $E_0=\Gamma_0 k_BT$.   Following~\cite{elfringJFM2016}, we define the dimensionless surface compressibility as the ratio of viscous over surface tension gradient forces
\begin{equation}
\beta = \frac{\eta V_0}{E_0}  \ .
\label{beta}
\end{equation}
At low injection speeds, the surfactant layer is hardly perturbed by the flow and the interface behaves as a solid wall ($\beta \to 0$).  Conversely, the elastic contribution of surfactants is irrelevant at high speeds. The fully compressible limit ($\beta \to \infty$) therefore coincides with the no-stress boundary condition for a perfectly clean interface. Note that the crossover value  $\beta_c \sim \mathcal{O}(1)$  that separates the two regimes can actually be reached for a very low surface density. Indeed,   for water  at room temperature and with $V_0=10^{-3}$~m$\cdot$s$^{-1}$, the value $\beta_c =1$ therefore corresponds to a surface concentration as low as  $\Gamma_{0}\approx 250$~molecules$\cdot \mu$m$^{-2}$. Interestingly, this estimate is very similar to that ($\approx 300$ molecules$\cdot \mu$m$^{-2}$) invoked by Hu and Larson to account for the suppression of  Marangoni flows in evaporating droplets~\cite{huLangmuir2005}. This shows that a minute amount of surfactants is sufficient to strongly affect the overall structure of the flow.

\section{Analytical description of a confined Landau-Squire jet}
\label{analytical}

The general problem defined by Eqs.~(\ref{stokes})--(\ref{elaslangmuir})  is nonlinear and far too complex to be tractable analytically.
Some simplifications are then needed in order to be predictive. First, we focus the discussion on the stationary regime. Although time-dependent behaviors might be relevant in pressure-relaxation experiments for instance~\cite{peaudecerfPNAS2017}, they are not considered here. Second, we make the hypothesis that the surface concentration is sufficiently small so that the nonlinear contributions to the Gibbs elasticity are irrelevant. The Marangoni boundary condition~(\ref{bcMar}) is then expressed as
\begin{equation}
 \partial_z v_r \big\vert_{z=0} =  -\frac{E_0}{\eta \Gamma_0} \frac{\partial \Gamma}{\partial r}  \ .
\label{mbc}
\end{equation}
Third, we concentrate on length scales that are much larger than the injection radius~$a$, but at the same time much smaller than the cell size~$R$. In the discussion that follows, we therefore consider that the system is unbounded in the horizontal directions. The velocity and concentration fields are then expected to relax to their unperturbed values
\begin{equation}
\lim_{\vert \mathbf{r} \vert  \to \infty} \mathbf{v}(\mathbf{r}) = \bm{0} \ , \quad \text{and} \quad \lim_{r  \to \infty} \Gamma(r) = \Gamma_0 \ . 
\label{bcbulk}
\end{equation}
It is also assumed that the depth of the container is infinitely large, so that $H$ is the only relevant length scale in the vertical direction.

\subsection{Landau-Squire jet near a clean interface}
\label{landausquire}

We consider the flow emerging from a narrow tube in a quiescent liquid. The tube is vertical and oriented upward, with its extremity that lies at~$\mathbf{r}_0=(0,0,-H)$ (see Fig.~\ref{fig1}).
Our analysis is based on an original solution proposed by Landau~\cite{landaubook} and Squire~\cite{squireQJMAM1951}, which has proven adequate to describe small-scale fluidic jets~\cite{laoNanoLett2013, secchiNature2016,secchiJFM2017}.   
Let us denote  $V_{\mathrm{inj}}$  the average velocity across the section of the jet.
The mass and momentum fluxes are respectively given by $Q=\rho \pi a^2 V_{\mathrm{inj}}$ and $P=\rho \pi a^2 V_{\mathrm{inj}}^2$. If we now take the limit $a\to 0$ but  keeping  the  momentum flux $P$ to a fixed value, the flow can then be regarded as originating from a point source located at $\mathbf{r}_0$. Momentum transfer at the opening of the narrow tube might therefore be approximated by a stokeslet of strength $P$ (which has the dimension of a force). The latter is a fundamental solution of the Stokes equations in a viscous fluid under the action of the force density $\mathbf{f}(\mathbf{r}) = P  \delta (\mathbf{r}-\mathbf{r}_0) \mathbf{e}_z$. This singularity produces the flow field $\mathbf{v} (\mathbf{r}) =  \mathsf{G}(\mathbf{r}-\mathbf{r}_0) \cdot P \mathbf{e}_z$, where the tensor $\mathsf{G}=\left(\mathbb{I}r^2 - \mathbf{rr}\right)/(8\pi \eta r^3)$ is the free space Green's function ($\mathbb{I}$ being the identity tensor).
A peculiar feature of the Landau-Squire flow is that the mass flux $Q=(\rho \pi a^2 P)^{1/2}\to 0$ actually vanishes  in the limit $a \to 0$ at fixed $P$~\cite{laoNanoLett2013}. The flow  is thus entirely determined by the transfer of momentum from the jet to the surrounding liquid~\cite{landaubook}.

In the vicinity of the interface, however, the Landau-Squire solution has to be modified in order to enforce the boundary condition~(\ref{bcint}). Using the method of images, the solution~$\mathbf{v}^{(0)}$ of the Stokes equations for a clean interface (\textit{i.e.},  in the absence of surfactants) is readily obtained  as~\cite{ekielEPJE2018}
\begin{equation}
\mathbf{v}^{(0)} (\mathbf{r}) = \left[ \mathsf{G}(\mathbf{r}-\mathbf{r}_0) - \mathsf{G}(\mathbf{r}+\mathbf{r}_0) \right] \cdot P \mathbf{e}_z \ .
\label{freeint}
\end{equation}
The fictitious singularity located at $\mathbf{r}_0'=-\mathbf{r}_0$ is the image of the point force that ensures the boundary condition~(\ref{bcint}).
The components of the velocity field are then expressed in  cylindrical coordinates as
\begin{subequations} \label{v0}
\begin{align}
& v_r^{(0)} (r,z) =\frac{V_0}{4}rH  \left[ \frac{(z+H)}{r_+^3} - \frac{(z-H)}{r_-^3} \right] \ , \label{vr0} \\
& v_z^{(0)} (r,z) =\frac{V_0}{4}H  \left[\frac{1}{r_+}- \frac{1}{r_-} +\frac{(z+H)^2}{r_+^3} - \frac{(z-H)^2}{r_-^3} \right] \ , \label{vz0}
\end{align}
\end{subequations}
with $r_{\pm} = \sqrt{ r^2+(z\pm H)^2 }$, and where we set $V_0=P/(2\pi \eta H)$.

\begin{figure}
\centering
\includegraphics[width=0.9\columnwidth]{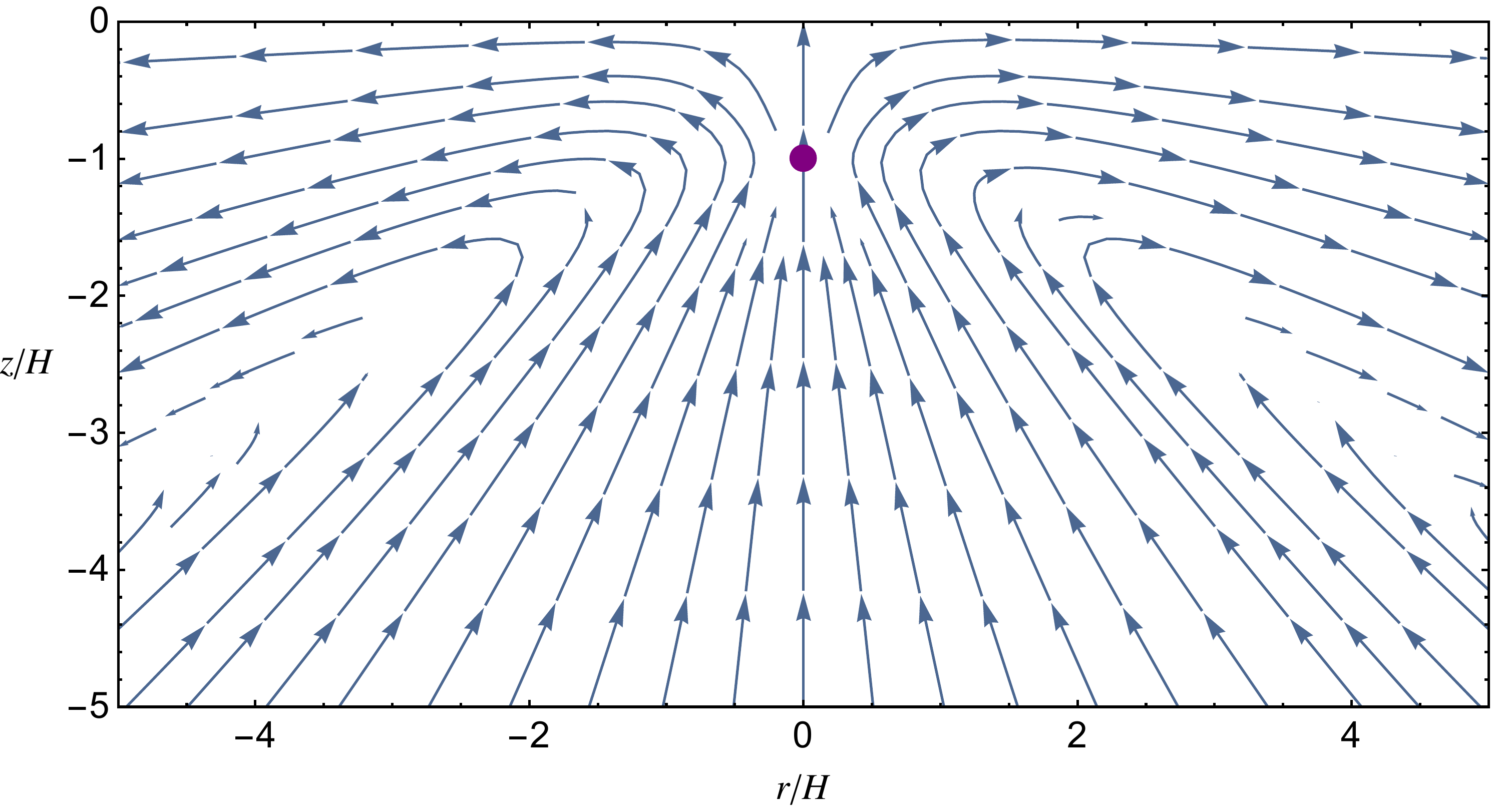}
\caption{Streamlines of the Landau-Squire jet perpendicular to a clean interface [see Eq.~(\ref{v0})]. The point source is marked by a purple dot.}
\label{fig2}
\end{figure}

The  streamlines corresponding to the flow  Eqs.~(\ref{v0}) are plotted in Fig.~\ref{fig2}. The  morphology of the flow is that of an open torus. Indeed, since the liquid domain is unbounded, the streamlines are expected to close at infinity.
The velocity scale $V_0$ defined above is related to the maximum  velocity of the flow along the interface: $v_{r,\mathrm{max}}(r,0)\propto V_0$.
At this point, it is important to note that $V_0$ actually differs from the injection velocity~$V_{\mathrm{inj}}$. 
As a matter of fact, the limit of low Reynolds number  implies a proportionality relation between $V_0$ and $V_{\mathrm{inj}}$. This  issue will be discussed in connection with the simulations in  Sec.~\ref{numerical}.


\subsection{Surfactant-laden interface}

We now consider an interface covered with insoluble, surface-active molecules. Starting from an initial homogeneous distribution, surfactant molecules are first advected by the flow. The resulting concentration gradient then exerts a shear stress on the fluid, which in turn modifies the hydrodynamics. We discuss here the stationary limit of this process, in the regime $\text{Pe} \gg 1$. The diffusion term is therefore disregarded in the transport~Eq.~(\ref{adeq}), which can then  be integrated once to give
\begin{equation}
 v_r(r,0)  \Gamma (r) = 0 \ .
 \label{adcond}
\end{equation}
This condition is reminiscent of the stagnant cap condition first proposed by Levich in the context of the buoyant motion of a bubble rising in a liquid~\cite{levichbook,palaJFM2006}.

The mathematical model defined by Eqs.~(\ref{stokes}) and~(\ref{adcond}), together with the boundary conditions (\ref{bcint}), (\ref{mbc}) and~(\ref{bcbulk}),  describes the rearrangement of surfactants in response to the incident jet flow~$\mathbf{v}^{(0)}$.
To solve the hydrodynamic problem, we decompose the total velocity field as $\mathbf{v}(\mathbf{r})=\mathbf{v}^{(0)}(\mathbf{r})+\mathbf{v}^{(1)}(\mathbf{r})$, where $\mathbf{v}^{(1)}$ is sought as a regular solution of the Stokes equations.
This is conveniently achieved in 2D Fourier representation. 
The problem being radially symmetric, we introduce the Hankel transforms of order $\nu$~\cite{piessensbook}
\begin{subequations}
\begin{align}
&  \tilde{f}(q,z) = \mathcal{H}_{\nu} \big[f(r,z) \big] = \int_0^{\infty} r J_{\nu} (qr) f(r,z)   \d r    \ ,  \\
&  f(r,z) = \mathcal{H}^{-1}_{\nu} \big[\tilde{f}(q,z) \big] =  \int_0^{\infty} q J_{\nu} (qr) \tilde{f}(q,z)   \d q    \ .  
\end{align}
\end{subequations}
Define  $\tilde{v}^{(1)}_z (q,z)= \mathcal{H}_0 \big[v_z^{(1)} \big]$  and
$\tilde{v}^{(1)}_r (q,z)= \mathcal{H}_1 \big[v_r^{(1)} \big]$,  
it can be shown that the flow equations~(\ref{stokes}) in the liquid phase ($z<0$) assume the following form~\cite{tanzoshJFM1995,bickelPRE2007}
\begin{subequations} \label{stokeshankel}
\begin{align}
& \frac{\partial^4 \tilde{v}^{(1)}_z}{ \partial z^4} -2q^2 \frac{\partial^2 \tilde{v}^{(1)}_z}{ \partial z^2}+ q^4  \tilde{v}^{(1)}_z = 0 \ , \\
& q  \tilde{v}^{(1)}_r + \frac{ \partial  \tilde{v}^{(1)}_z}{\partial z} =0 \ .
\end{align}
\end{subequations}
The solution  that satisfies both the boundary condition $\tilde{v}^{(1)}_z (q,0)=0$ while vanishing far away from the  interface, $\lim_{z\to -\infty} \tilde{v}^{(1)}_z (q,z)=0$,   then reads
\begin{subequations} \label{v1hankel}
\begin{align}
& \tilde{v}^{(1)}_z (q,z) = A(q) z e^{qz} \ , \label{v1zhankel} \\
& \tilde{v}^{(1)}_r(q,z)  = - q^{-1} A(q) (1+qz) e^{qz} \ , \label{v1rhankel}
\end{align}
\end{subequations}
where the integration constant $A(q)$ remains yet to be determined.

Eq.~(\ref{v1hankel}) is the general solution of the Stokes problem. To specify the Marangoni counterflow, we now express the excess density~$\delta\Gamma \doteq \Gamma - \Gamma_0$ thanks to the stress continuity condition.
Taking the Hankel transform of Eq.~(\ref{bcMar}) together with the definition $\delta \tilde{\Gamma} (q)=\mathcal{H}_0 \left[ \delta \Gamma  \right]$,
one finally arrives at 
\begin{equation}
 \delta \tilde{\Gamma} (q) = - \frac{2\eta \Gamma_0}{E_0} q^{-1}A(q) \ .
 \label{gammaq}
\end{equation}
With this relation, we have  established the general solution of the coupled transport problem.
However, we still need to account for the closure relation  Eq.~(\ref{adcond}) in order to determine $A(q)$. We shall see in the following sections that this nonlinear problem admits two analytical solutions,
depending on the surface compressibility regime.

\subsubsection{Low-compressibility regime}

\begin{figure}
\centering
\includegraphics[width=0.9\columnwidth]{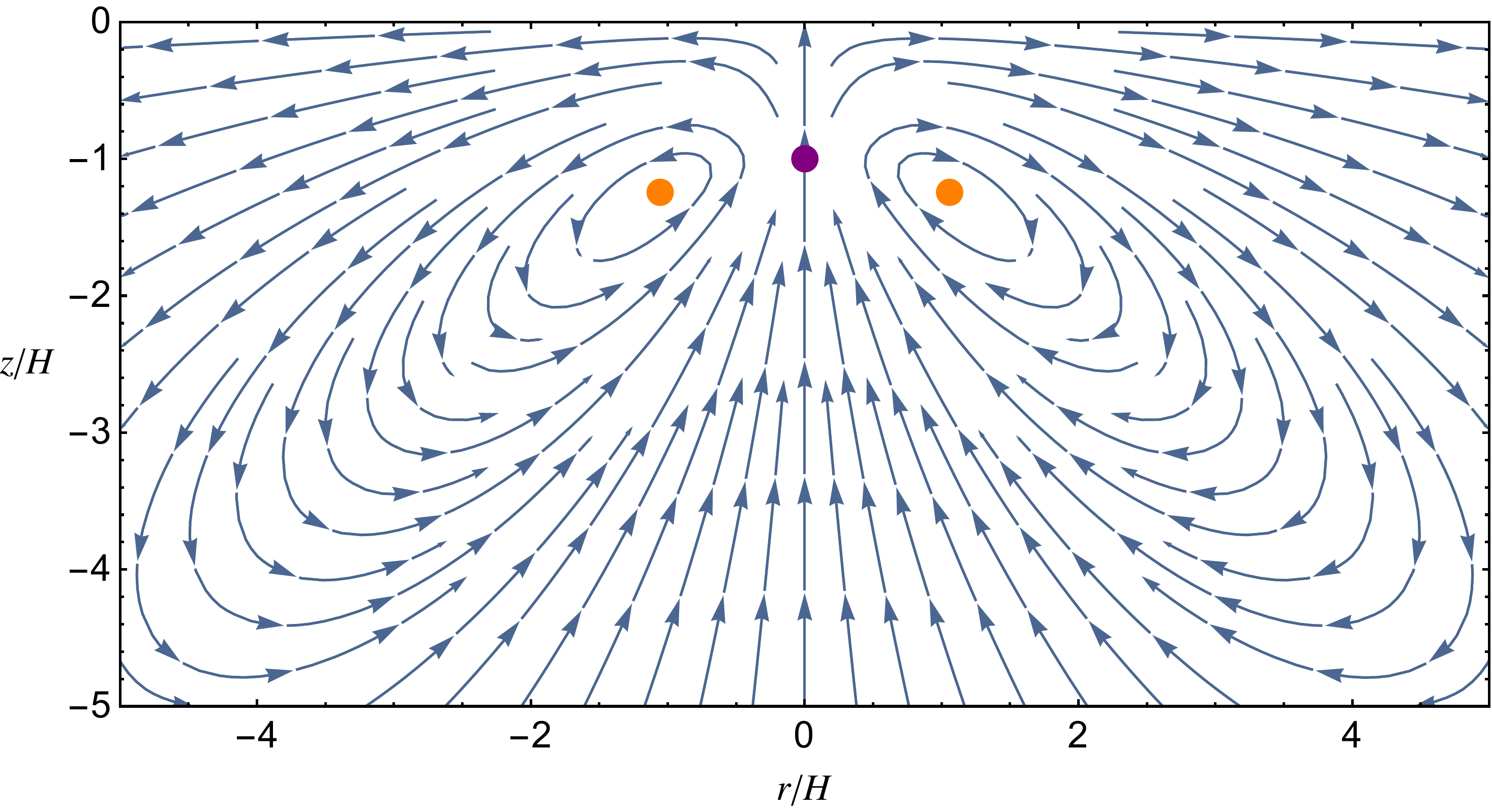}
\caption{Streamlines of the total velocity field $\mathbf{v}=\mathbf{v}^{(0)}+\mathbf{v}^{(1)}$ in the low-compressibility regime $0<\beta<1$. The point source is marked by a purple dot. The orange dots indicate the position of the centerline of the vortex ring.}
\label{fig3}
\end{figure}

Eq.~(\ref{adcond}) states that the product $ v_r(r,0)  \Gamma (r)$ vanishes all along the interface. Let us first assume that the surface compressibility~$\beta$ is sufficiently small so that
the concentration remains finite everywhere. Doing so, Eq.~(\ref{adcond}) finally comes down to
\begin{equation}
 v_r(r,0) = 0 \ .
\end{equation}
At low~$\beta$, the interface appears so rigid that it remains perfectly still, even though the liquid is not quiescent in the bulk.
This absence of motion results from the exact cancellation
of the base flow $\mathbf{v}^{(0)}$ and the Marangoni counterflow $\mathbf{v}^{(1)}$  at $z=0$.
But since $\mathbf{v}^{(0)}$ is given by Eq.~(\ref{v0}), it is therefore  straightforward to get
\begin{equation}
A(q) = \frac{V_0 H^2}{2} q e^{-qH} \ .
\label{aqlow}
\end{equation}
Taking the inverse Hankel transform then leads to the velocity components of the counterflow
\begin{subequations} \label{v1}
\begin{align}
& v_r^{(1)} (r,z) = - \frac{V_0}{2}H^2   \left[r^2+(H+2z)(H-z)\right] \frac{r}{r_-^5}   \ , \label{vr1} \\
& v_z^{(1)} (r,z) = \frac{V_0}{2}H^2   \left[2(H-z)^2-r^2\right] \frac{z}{r_-^5} \ . \label{vz1}
\end{align}
\end{subequations}
The streamlines corresponding to  the total velocity field $\mathbf{v}=\mathbf{v}^{(0)}+\mathbf{v}^{(1)}$
are plotted in Fig.~\ref{fig3}. The morphology of the flow is again toroidal, but the striking feature when comparing to Fig.~\ref{fig2} is that the streamlines now close
at some finite distance from the injection point. The presence of insoluble surfactants thus strongly modifies the 3D structure of the flow.
The position~$(R_{\mathrm{torus}},Z_{\mathrm{torus}})$ of the centerline of the vortex is determined numerically: we find $R_{\mathrm{torus}} \approx 1.056H$ and $Z_{\mathrm{torus}} \approx 1.248H$. Interestingly, it can be noticed that the morphology of the flow does not depend on the velocity scale $V_0$.

\begin{figure}
\centering
\includegraphics[width=0.9\columnwidth]{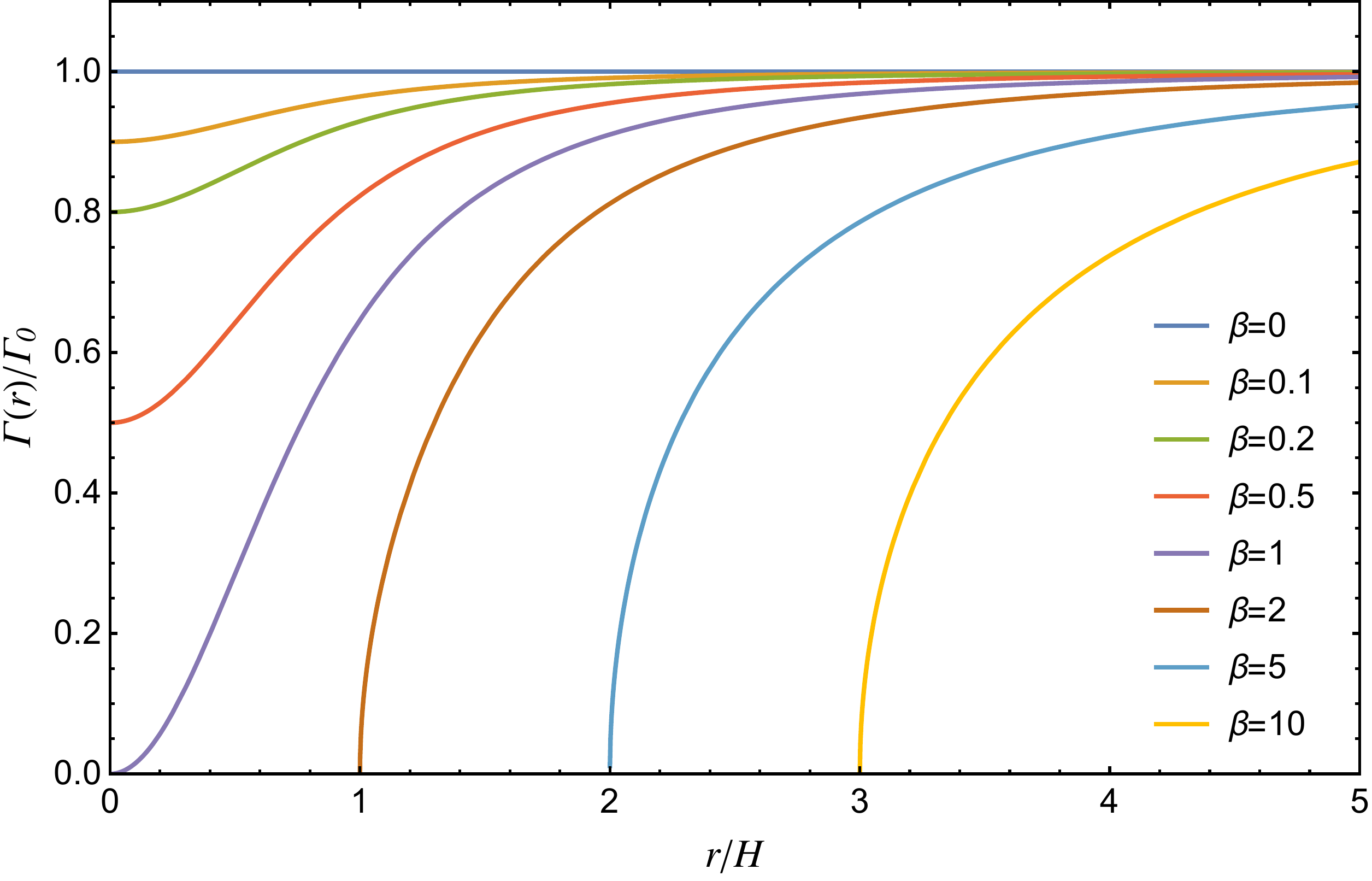}
\caption{Evolution of the surfactant concentration~$\Gamma(r)$ with the compressibility~$\beta=\eta V_0/E_0$.  The curves correspond to Eq.~(\ref{gamma0b1}) for $0\leq \beta \leq 1$; for $\beta>1$, the concentration is obtained by numerical inversion of its Hankel transform [Eqs.~(\ref{gammaq}) and~(\ref{aqhigh})].   Note the cross-over between the low- and high-compressibility regimes  that occurs when $\beta=1$.}
\label{fig4}
\end{figure}

The concentration field is then deduced from Eq.~(\ref{gammaq}).
After inversion of the Hankel transform, we obtain the  expression
\begin{equation}
\Gamma (r)  = \Gamma_0 \left[ 1-\beta \frac{H^3}{(r^2+H^2)^{3/2}} \right]  \ .
\label{gamma0b1}
\end{equation}
This distribution  is plotted in Fig.~\ref{fig4} for different values of the compressibility. As $\beta$ increases, the concentration at the origin decreases as $\Gamma(0)=\Gamma_0 (1-\beta)$. But obviously this solution ceases to be valid when $\beta=1$. The low-compressibility regime is therefore restricted to the range $0 < \beta < 1$.

\subsubsection{High-compressibility regime}

The situation gets more involved when $\beta >1$. Coming back to Eq.~(\ref{adcond}), we now state that either the velocity or the concentration vanishes at some place or another along the interface. From the physics viewpoint, we are led to assume the existence of a critical radius~$r_c$ that separates two regions such that
\begin{subequations}
\label{condbc}
\begin{align}
& \Gamma (r) = 0 \ ,  \qquad 0\leq r <r_c \ , \label{concinf} \\
& v_r(r,0) =0 \ ,  \qquad r > r_c   \ .  \label{vsup}
\end{align}
\end{subequations}
The first relation expresses that surfactant molecules are entirely depleted from the inner region $r<r_c$. In the outer region $r>r_c$, the  counterflow exactly cancels the base flow so that the total velocity vanishes at the interface, as discussed in the previous section.

The mixed boundary value problem defined by Eqs.~(\ref{concinf}) and~(\ref{vsup}) can be then recast in terms of integral equations. Indeed, taking the inverse Hankel transform of Eqs.~(\ref{v1rhankel}) and~(\ref{gammaq}) and using the identity $\partial_r\left[ r J_1(qr) \right] = qr J_0(qr)$, one arrives at the equivalent  set of dual integral equations
\begin{subequations}
\label{dualbc}
\begin{align}
& \int_0^{\infty} A(q) J_0(qr)  \d q = \frac{E_0}{2\eta} \ , \qquad 0\leq r <r_c \ , \label{dualbc1} \\
& \int_0^{\infty}  qA(q) J_0(qr)\d q   =\frac{1}{r}  \left[ r f(r) \right]' \ , \qquad r > r_c   \ ,  \label{dualbc2}
\end{align}
\end{subequations}
with $f(r) = V_0H^2r/[2(r^2+H^2)^{3/2}]$. The general idea to solve dual integral equations is to build up a solution in such a way that part of the problem is automatically satisfied~\cite{sneddonbook,duffybook}.  The ensuing derivation being quite technical but not essential for the argumentation, we refer the interested reader to Appendix~\ref{appAq} for a detailed account of the calculations. We then find after some algebra that~$A(q)$ is given by
\begin{equation}
A(q) = \frac{V_0 H^2}{2} q e^{-qH} - \frac{2 V_0}{  \pi  q } \int_{0}^{r_c/H} \frac{xF(x,qH)}{(1+x^2)^2}   \d x \ ,
\label{aqhigh}
\end{equation}
with $F(x,y) = \sin (xy) -xy \cos (xy)$. The first term on the right-hand-side has already been derived in the low-compressibility regime, see Eq.~(\ref{aqlow}). The second  contribution exists only for $r_c>0$. In the limit $r_c \to \infty$, the integral can be calculated and is found to  cancel exactly the first term: when the size of the depletion zone is infinitely large, the concentration is zero everywhere and one recovers the clean interface limit.

\begin{figure}
\centering
\includegraphics[width=0.9\columnwidth]{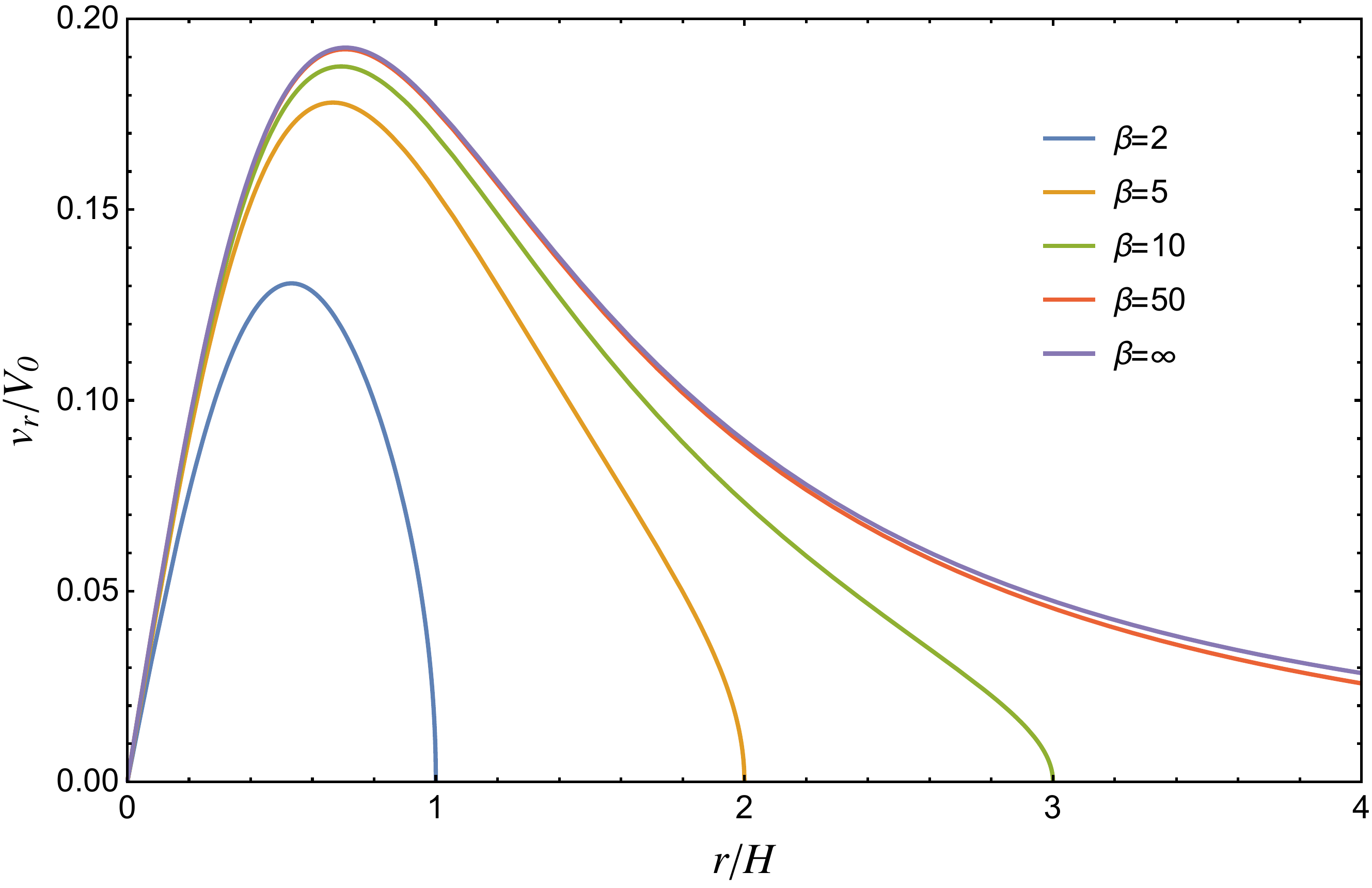}
\caption{Interfacial velocity field $v_r(r,0)$ in the high-compressibility regime $\beta >1$. The limit $\beta \to \infty$ corresponds to the clean interface limit.}
\label{fig5}
\end{figure}

At this point, we still have to figure out the (yet unknown) radius of the depletion zone. Since the surfactants are insoluble,  conservation of the total number of molecules leads to the relation
\begin{equation}
\int_0^R 2 \pi r \Gamma (r)  \d r = \pi R^2 \Gamma_0 \ ,
\end{equation}
with $R$ the size of the system~\cite{remark}. In the limit $r_c, H \ll R$, the integrals can be evaluated exactly  and we finally arrive at
\begin{equation}
r_c = H \sqrt{\beta - 1} \ .
\end{equation}
This relation shows that the sweeping mechanism of surfactants does not involve any new length scale.
Notice also that this solution  exists only in the high-compressibility regime $\beta >1$, as  expected from the previous discussion.

Having established the expressions of $A(q)$ and $r_c$, we can now compute the velocity and the concentration fields by inverting numerically the Hankel transforms in Eqs.~(\ref{v1hankel}) and~(\ref{gammaq}), respectively.
Still, it can be shown analytically that   $\Gamma(r) \sim \sqrt{r-r_c}$ when $r \to r_c^+$: the asymptotic behavior
 of $\Gamma(r)$ is thus singular in the vicinity of $r_c$. The distribution of surfactants is represented in Fig.~\ref{fig4}; we also plot the interfacial velocity in Fig.~\ref{fig5}.

\section{Numerical simulations}
\label{numerical}

\subsection{Simulation method}

In the previous section, we have managed to establish analytically the hydrodynamic response of a surfactant-laden interface to a stokeslet.  Still, the finite sizes of both the container and the injection tube have been neglected so far. Diffusion was not considered either, even though it may become relevant at the border of the depleted region,  where the concentration varies rapidly.  These effect are now accounted for through numerical simulations using a commercial finite element computational software (COMSOL Multiphysics$^\circledR$~\cite{comsol}). In the simulations, the liquid is injected into a cylindrical cell through a cylindrical pipette. 
The inlet flow field is normal to the pipette cross-section and has a uniform velocity~$V_{\mathrm{inj}}$. 
 An outflow (purge) pressure condition is set up at the peripheral bottom ridge of the container. All solid walls, including that of the pipette, feature a no-slip boundary condition. 
Technical details regarding the simulations are given in Appendix~\ref{appB}.

The governing equations are first made dimensionless by choosing the pipette radius~$a$ and the injection speed~$V_{\mathrm{inj}}$ as the length and velocity scales, respectively.  Two different container sizes are considered: a ``small'' cell  of radius~$R=72a$ and  a  ``large'' one with~$R=144a$. We keep  the same height~$L=120a$.  In accordance with the theoretical Sec.~\ref{model}, the simulations probe the regime of high P\'eclet number $\text{Pe}_{\mathrm{num}}=aV_{\mathrm{inj}}/D$. Hereafter, the value of the P\'eclet number is arbitrarily set to  $\text{Pe}_{\mathrm{num}}=3.57\times 10^3$. 

In the simulations, the concentration of insoluble surface-active molecules dispersed on the water-air interface is controlled by the  fraction $x=\Gamma_0/\Gamma_{\infty}$ of area covered with surfactants. This parameter is varied from $x\sim 10^{-4}$, which is extremely dilute  (about 230 molecules$\cdot \mu$m$^{-2}$), up to $x\sim 10^{-1}$, which corresponds to a moderate coverage. Higher surface coverages are not considered here since we focus on the regimes where only traces of surfactants are present. The dimensionless compressibility is then defined as $\beta_{\textrm{num}} = \eta V_{\mathrm{inj}}/(\Gamma_0 k_B T)$. We also introduce the quantity $\beta_{\textrm{num}}^{\infty} = \eta V_{\mathrm{inj}}/(\Gamma_{\infty} k_B T)$, so that we have the relation $x=\beta_{\textrm{num}}^{\infty}/\beta_{\textrm{num}}$. In this work, the value of $\beta_{\textrm{num}}^{\infty}$ is arbitrarily set to $\beta_{\textrm{num}}^{\infty}=1.06 \times 10^{-3}$. Note however that the Gibbs elasticity coefficient is not constant in the simulations, but is given by the full nonlinear expression~(\ref{elaslangmuir}).

\subsection{Surfactant-free situation}

\begin{figure*}
\includegraphics[width=12cm]{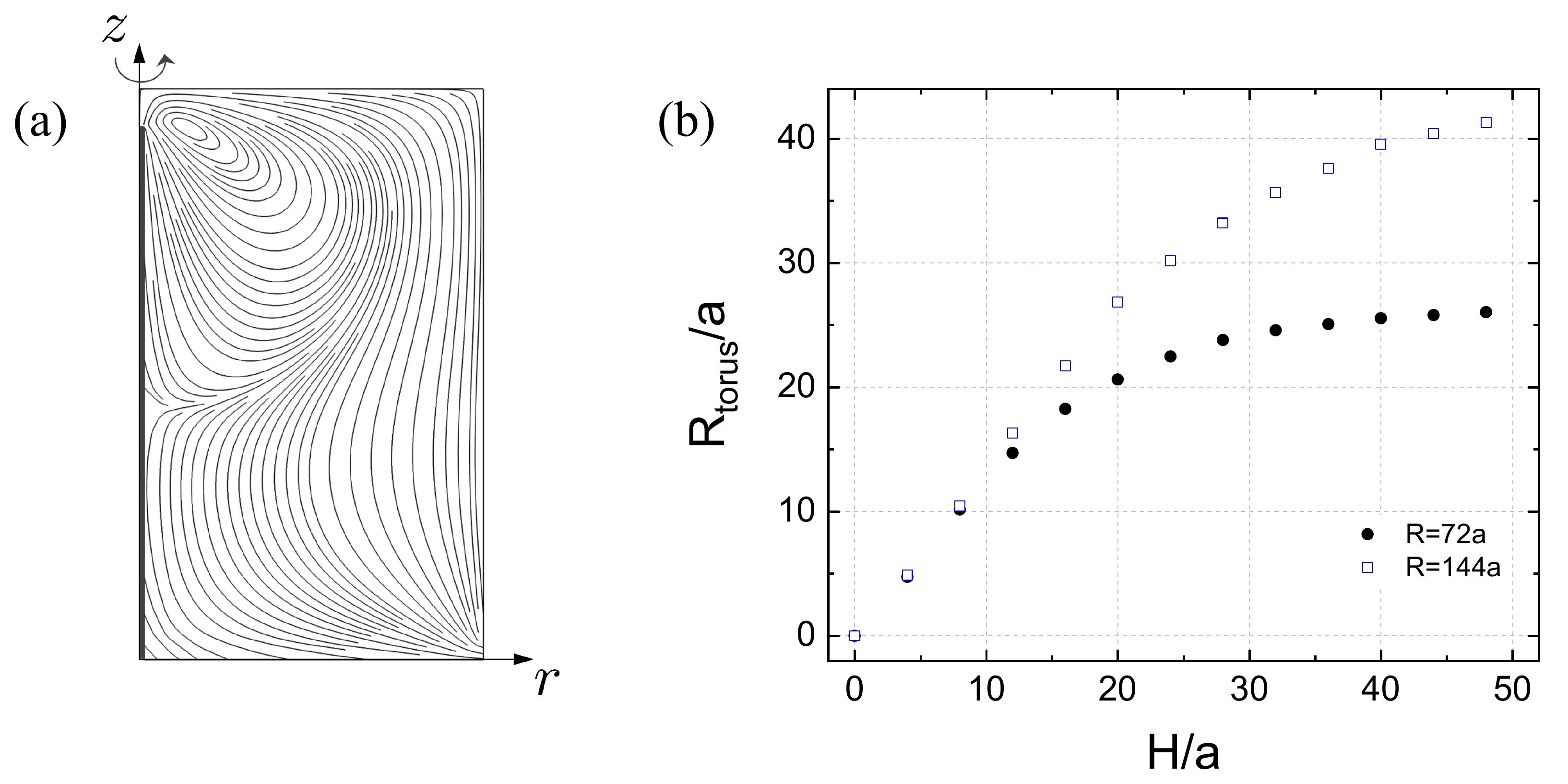}
\caption{(a) Flow streamlines for  a clean interface ($x=0$). Small cell ($R=72a$),  gap $H=8a$. (b)~Radial position of the  torus centerline for  a clean interface ($x=0$).   }
\label{fig6}
\end{figure*}

We first carry out the simulations in the idealized case of a pure interface devoid of any kind of surface-active species. Fig.~\ref{fig6}(a) shows the flow streamlines in the small cell for a gap $H=8a$. The structure of the flow is toroidal, as expected. But unlike the results obtained in Sec.~\ref{landausquire},  the streamlines are now closed due to the finite size of the system. The evolution of the radial position of the centerline, $R_{\mathrm{torus}}$, as a function of the gap~$H$ is shown in Fig.~\ref{fig6}(b)  for the two different cell sizes. For $0 <H \lesssim 12 a$, $R_{\mathrm{torus}}$ first increases linearly with~$H$, which is actually the relevant length scale at intermediate gap $a \ll H \ll R$.  As the gap increases further, the position of the centerline tends to saturate to a value that is controlled by the size of the system.

\begin{figure*}
\includegraphics[width=13cm]{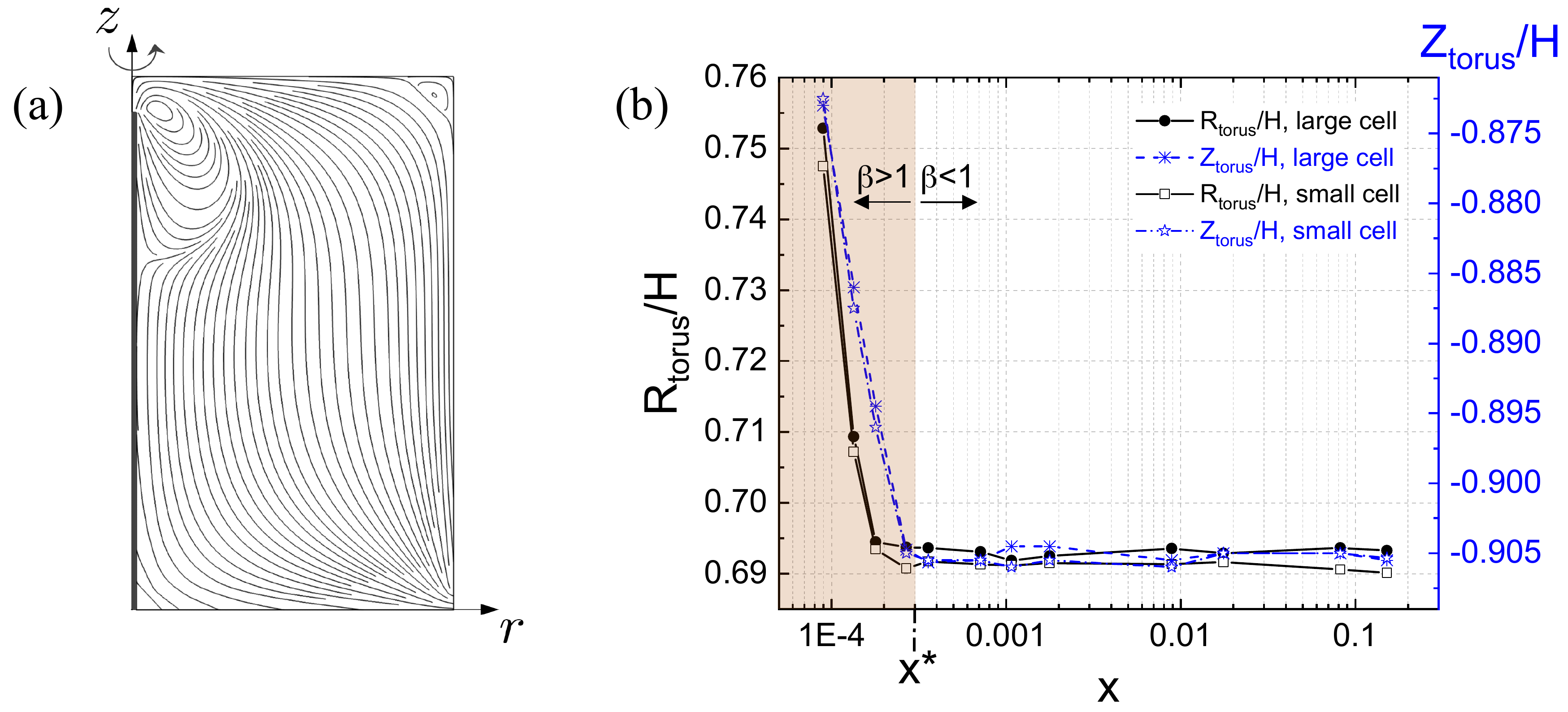}
\caption{(a) Flow streamlines  for  a surfactant-laden interface ($x=0.152\,$). Small cell ($R=72a$), gap $H=8a$. (b) Evolution of the position of the  torus centerline. Large cell ($R=144a$) and small cell ($R=72a$), gap $H=8a$. The shaded area corresponds to the high-compressibility regime.}
\label{fig7}
\end{figure*}

The surfactant-free situation may also serve as a reference state in order to relate the parameters of the simulations --- namely, the size~$a$ of the injection nozzle and the injection velocity~$V_{\mathrm{inj}}$ --- to those of the analytical theory. As a matter of fact, only~$V_0$ matters in the latter case. At low Reynolds number, both velocity scales $V_0$ and $V_{\textrm{inj}}$ must be proportional to each other. We thus define the  proportionality factor $h=V_0/(2V_{\textrm{inj}})$, which is a function of the gap $H/a$.
As explained in App.~\ref{appC}, we can extract $h(H/a)$  from the slope of the interfacial velocity field $v_r(r,0)$ in the vicinity of the origin. Our numerical results are consistent with a power-law  behavior
\begin{equation}
h\left( x \right)  = K x^{-\alpha}   \ ,
 \label{powerlaw}
\end{equation}
with $K$ and $\alpha$ two fitting parameters whose numerical values are: $\alpha \approx 1.38$ and $K \approx 1.60$.
We find in particular that $V_0 \to 0$ as $H\to \infty$ at fixed $V_{\textrm{inj}}$, as expected.

\subsection{Surfactant-laden interface}

In the presence of surfactants, the global structure of the flow  may at first sight seem similar to the surfactant-free situation. This is illustrated by the streamlines plotted on Fig.~\ref{fig7}(a) for $x=0.152$ and $H=8a$. Still, it can be noticed that the radial extension of the torus is definitely smaller when surfactants are present. In addition, a secondary centripetal roll appears below the free surface at the periphery of the cell. This feature is also a signature of the presence of surfactants.

\begin{figure*}
\includegraphics[width=13cm]{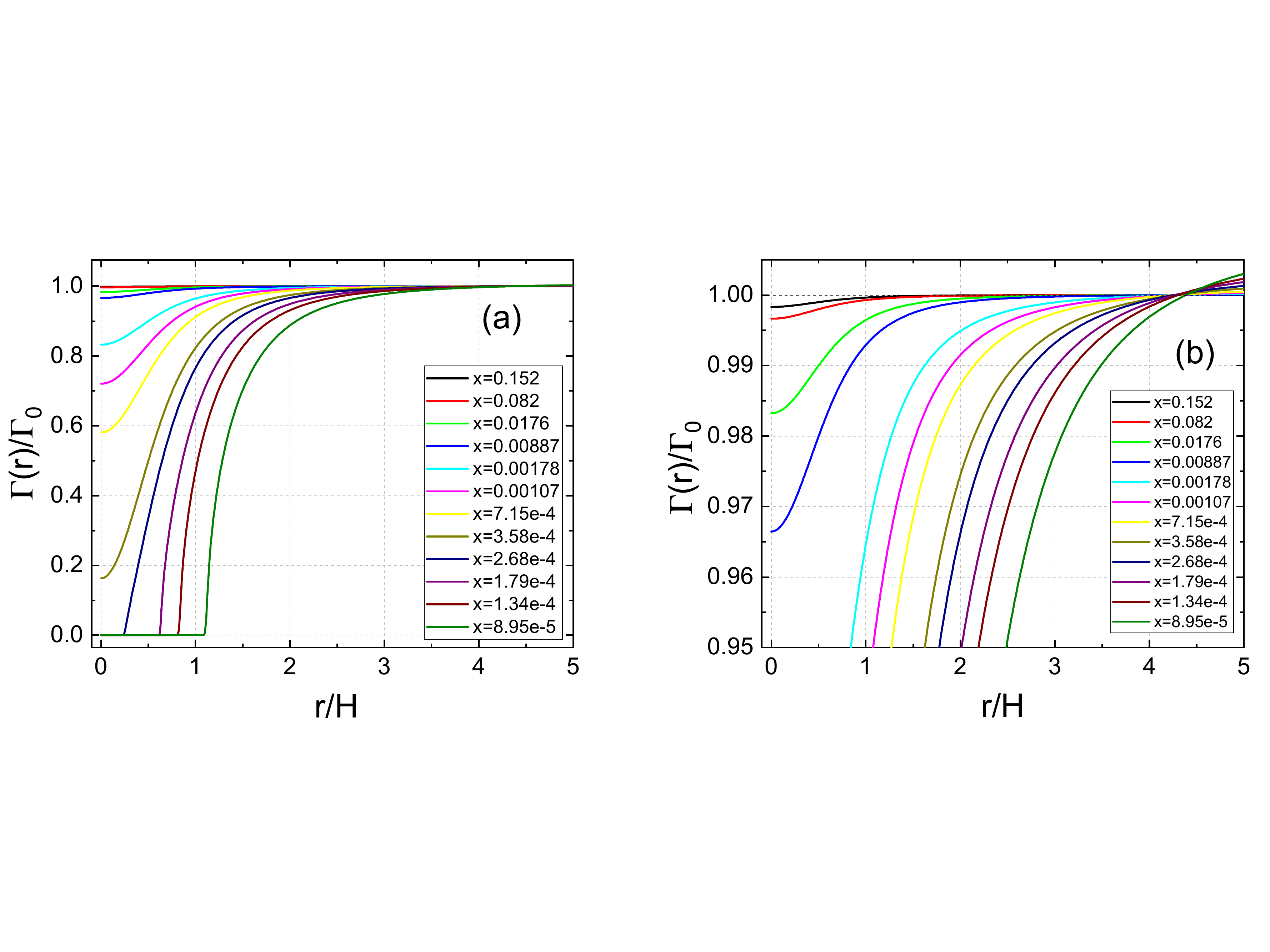}
\caption{(a) Evolution of the concentration of surfactant  for different surface coverages. Large cell ($R=144a$), gap $H=8a$. (b) Same as (a) but the graph is zoomed-in to better visualize the concentration profiles for the largest values of $x$.}
\label{fig8}
\end{figure*}

\begin{figure*}
\includegraphics[width=13cm]{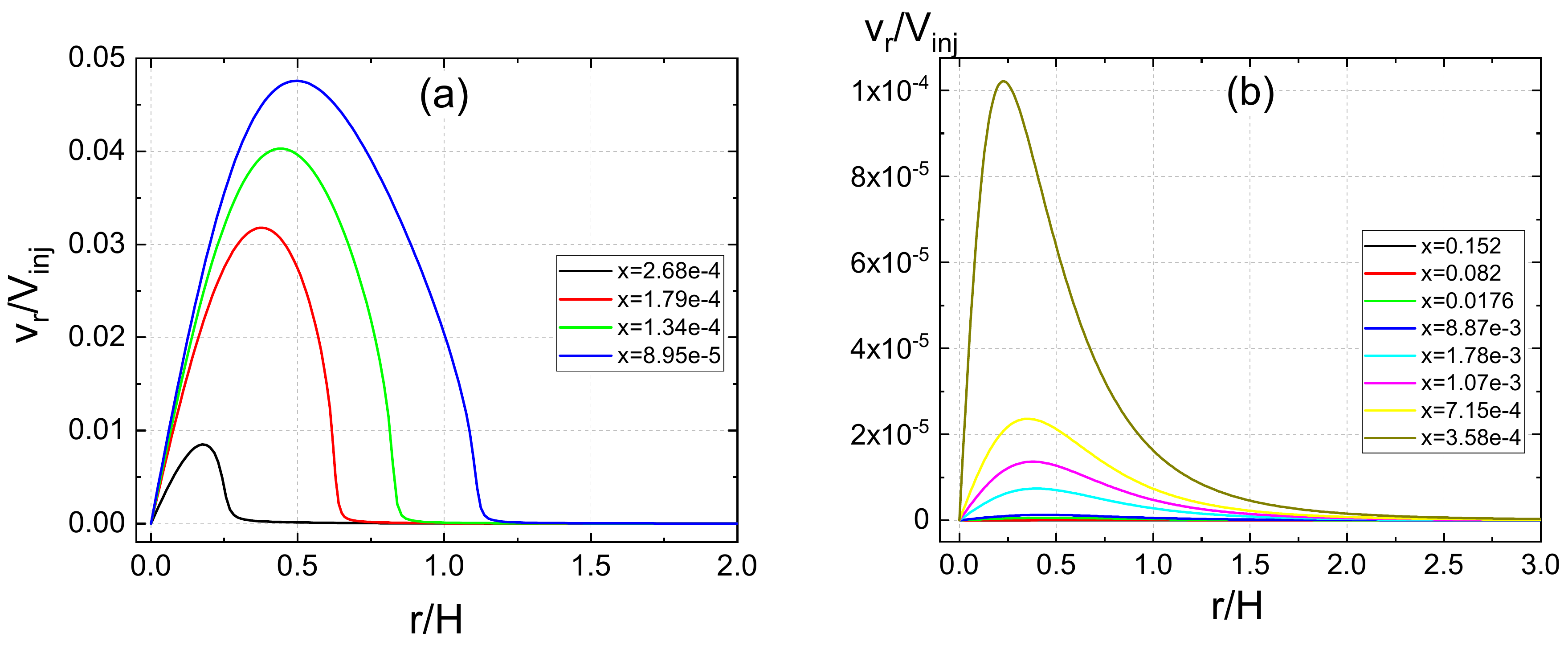}
\caption{Evolution of the interfacial radial velocity profiles  for different surface coverages. Large cell ($R=144a$), gap $H=8a$. (a) High-compressibility regime ($x<x^\ast$). (b) Low-compressibility regime ($x>x^\ast$). Notice that the velocity scales are different between the two graphs.}
\label{fig9}
\end{figure*}

The simulations actually confirm that the toroidal structure is very sensitive to the presence of a minute amount of surfactants. For instance,  for $H=8a$, the radius of the torus exhibits a dramatic drop by almost 60\% between the clean interface situation ($R_{\mathrm{torus}}\approx 9.84a$ for $x=0$) and the smallest coverage value investigated in this work ($R_{\mathrm{torus}}\approx 6.04a$ for $x=8.95 \times 10^{-5}$). This trend is confirmed in Fig.~\ref{fig7}(b), which shows that the radial extension and the vertical position of the centerline decrease very rapidly in the high-compressibility regime ($x<x^\ast$) of the surfactant monolayer. But as soon as the low-compressibility regime is reached ($x>x^\ast$), both $R_{\mathrm{torus}}$ and $Z_{\mathrm{torus}}$ remain constant over almost 3 decades in surface concentration.
We can assess from the simulations that $x^\ast\approx 3.10^{-4}$, which corresponds to an initial surface concentration $\Gamma_0^\ast\approx 700$~molecules$\cdot \mu$m$^{-2}$.

Fig.~\ref{fig8} then shows that the surfactant molecules, initially uniformly distributed, are swept away by the radial flow and forced to accumulate at some distance from the fluid injection area. The simulations reproduce very well the theoretical trends --- compare with Fig.~\ref{fig4}. Decreasing the surface coverage $x$ amplifies the sweeping mechanism, and below a threshold value $x^\ast\approx 3.10^{-4}$, a depleted surfactant zone eventually occurs in the concentration profile. It is interesting to note that the cross-over value $x^\ast$, as determined from the properties of the bulk flow (Fig.~\ref{fig7}(b)), perfectly correlates  with that obtained from the features of the surface concentration (Fig.~\ref{fig8}(a)).

The corresponding interfacial velocities are graphed in Fig.~\ref{fig9}. In the high-compressibility regime $x<x^{\ast}$ [Fig.~\ref{fig9}(a)], one recovers the typical interfacial velocity profiles that vanish in the outer region, as predicted by the theory. But as soon as the low-compressibility regime is entered [Fig.~\ref{fig9}(b)],
the amplitude of $v_r(r,0)$ drops dramatically, even if the surfactant concentration is still very dilute. For instance, just above $x^\ast$ (for $x=3.58\times10^{-4}$), the maximum  velocity is already 2 orders of magnitude lower than that obtained with a clean interface. Further increasing the surface density makes the drop even more pronounced. The surfactant monolayer behaves essentially as a solid surface at the higher surface coverage probed in the simulations $x=0.152\,$.

\section{Conclusion}
\label{discussion}

To summarize, we have characterized the Stokesian hydrodynamic response of an interface to a radial flow in the presence of surface-active material. Emphasis was put on the dilute regime of surfactants. This study is thus   
complementary to the wealth of experimental and theoretical works that have been performed recently at high concentration in a similar geometry~\cite{rochePRL2014,rochePRE2016,bandiPRL2017,mandreJFM2017}. 
Here, we have shown that the presence of a minute amount of insoluble surfactants possesses a clear hydrodynamic signature. If the applied shear stress is lower than a critical value, the interface is motionless and behaves like a solid wall. Above the critical shear stress, the interface becomes partly mobile and the distribution of surfactants is singular at the border of the stagnant region.

These predictions are confirmed by numerical simulations. In particular, the transition between a low-compressibility and a high-compressibility regime clearly appears in Fig.~\ref{fig7}(b).  The cross-over occurs for the specific value $\beta_c=1$ of the dimensionless compressibility.
A quantitative comparison between theory and simulations can then be completed thanks to the proportionality relation between $V_0$ and $V_{\textrm{inj}}$.  Given the scaling form assumed by the proportionality factor Eq.~(\ref{powerlaw}), it is straightforward to get an estimate for the surface coverage $x^{*}\approx2.10^{-4}$. This value is in very good agreement with the simulation value $x^{*}\approx 3.10^{-4}$, all the more as there is an $\mathcal{O}\left( 10^{-4} \right)$ uncertainty in the determination of  $x^{*}$ (see Fig.~\ref{fig7}(b)).

Still, a puzzling issue lies in the difference between the predicted position of the vortex centerline and the simulation results in the low-compressibility regime. Although we find that both ratios are indeed independent of the surface coverage, the actual values differ between the theory ($R_{\mathrm{torus}} \approx 1.056H$ and $Z_{\mathrm{torus}} \approx 1.248H$) and the simulations ($R_{\mathrm{torus}} \approx 0.692H$ and $Z_{\mathrm{torus}} \approx 0.905H$) --- see Fig.~\ref{fig7}(b). This discrepancy is not related to the finite size of the container since we obtain the same limiting values for the small and the large cell. It might actually arise from the fact that the liquid is injected through a ``real'' tube in the simulations, whereas it is induced by a stokestlet in the theory. Clearly, both situations are not completely equivalent from a mathematical viewpoint, which could explain the discrepancy. Still, we emphasize that the invariant toroidal structure of the flow, which develops in the  low-compressibility regime with its centerline at a prescribed position, is a signature of the rigid boundary condition $v_r(r,0)=0$.

From an experimental viewpoint, the quantification of a small amount of surfactant dispersed at the water-air interface is an open issue~\cite{uematsuCOE2019}.
This requires to refine the theoretical models in order to provide reliable predictions regarding observable quantities. 
The results presented in this work is one attempt in this direction: we argue that the competition between the divergent outward flow and the solutal inward response may actually be used as a practical way to evidence the presence of impurities. In particular, the morphological features of the flow (\textit{i.e.}, the size and position of the torus) 
seem to be suitable candidates for a quantitative characterization of interfacial contamination.

Finally, let us mention that, although the presence of surfactants generally has a stabilizing effect~\cite{bergCES1965}, the situation is not always so clear. It was indeed suggested in a similar context that surface-active contaminants may actually induce the destabilization of the radial flow and lead to multipolar patterns~\cite{mizevPoF2005}.
The stability of the flow with respect to azimuthal perturbations would therefore deserve further investigation. 
Despite its apparent simplicity, the system under investigation might still reveal  a variety of unexpected features.

\begin{acknowledgments}

J-C. Loudet is indebted to the University of Bordeaux for financial support through the IdEx program ``D\'{e}veloppement des carri\`{e}res - Volet personnel de recherche''.

\end{acknowledgments}

\appendix

\section{Solution of the mixed boundary value problem}
\label{appAq}

In this appendix, we detail the general method leading to the solution of Eqs.~(\ref{dualbc1}) and~(\ref{dualbc2}).
Let us consider the  mixed boundary value problem defined by
\begin{subequations}
\label{eq0}
\begin{align}
& \int_0^{\infty}  A(q) J_0(qr) \d q = f_1(r) \ ,  & 0 \leq r<r_c \ , \label{eq01}\\
 &\int_0^{\infty}  qA(q) J_0(qr)\d q   = f_2(r)   \ ,  &  r\ > r_c \ . \label{eq02}
\end{align}
\end{subequations}
with $\left\{ f_1(r),f_2(r) \right\}$ a set of arbitrary functions.
Given the linearity of the equations, we can
assume the following decomposition
\begin{equation}
A(q)  = A_1(q) + A_2(q)
\end{equation}
where $A_1(q)$ and $A_2(q)$ satisfy Eq.~(\ref{eq0}) for the sets $\left\{ f_1(r),0 \right\}$ and $\left\{ 0,f_2(r) \right\}$, respectively. 
To proceed, we follow the general ideas  that consists in building up a solution in such a way that part of the problem is satisfied by construction  --- see for instance~\cite{sneddonbook,duffybook}.
The derivation involves the following integrals
\begin{equation}
\int_0^{\infty} J_0(qr) \cos (qt) \d q =
\begin{cases}
0 \quad \text{for} \quad 0 \leq r < t  \ , \\
(r^2-t^2)^{-1/2}  \quad \text{for} \quad r  > t  \ ,
\end{cases}
\label{intcos}
\end{equation}
as well as
\begin{equation}
\int_0^{\infty} J_0(qr) \sin (qt) \d q  =
\begin{cases}
(t^2-r^2)^{-1/2} \quad \text{for} \quad 0 \leq r < t  \ , \\
0  \quad \text{for} \quad r  > t  \ .
\end{cases}
\label{intsin}
\end{equation}

\subsection{Solution for the set $\left\{ f_1(r),0 \right\}$}

Let us first define an auxiliary function  $\Phi_1(t)$ such that
\begin{equation}
A_1(q)  \doteq \int_0^{r_c}  \Phi_1(t) \cos (qt) \d t \ , \label{defphi1}
\end{equation}
together with the condition $\Phi_1(0) = 0$. Integrating by parts gives
\begin{equation*}
A_1(q)  = q^{-1} \left( \Phi_1(r_c) \sin (qr_c) - \int_0^{r_c}  \Phi'_1(t) \sin (qt) \d t \right) \ ,
\end{equation*}
so that, thanks to (\ref{intsin}), the condition~(\ref{eq02}) is automatically satisfied.
To determine $\Phi_1(t)$, the definition~(\ref{defphi1}) is then inserted in Eq.~(\ref{eq01}). We thus get for $0 \leq r < r_c$
\begin{align*}
f_1(r)  & = \int_0^{r_c} \d t \, \Phi_1(t) \int_0^{\infty}   J_0(qr) \cos (qt) \d q   \\
&  = \int_0^{r}  \frac{\Phi_1(t)}{\sqrt{r^2-t^2}}   \d t  \ .
\end{align*}
The resulting Abel-type equation is readily inverted  and one obtains for $0 \leq t <r_c$
\begin{equation}
\Phi_1(t) = \frac{2}{\pi} \frac{\d}{\d t} \int_0^t \frac{rf_1(r)}{\sqrt{t^2-r^2}} \d r \ .
\label{resphi1}
\end{equation}

\subsection{Solution for the set $\left\{ 0,f_2(r) \right\}$}

We follow the same scheme for $A_2(q)$. Setting
\begin{equation}
A_2(q)  \doteq \int_{r_c}^{\infty}  \Phi_2(t) \cos (qt) \d t \ , \label{defphi2}
\end{equation}
with $\lim_{t \to \infty }\Phi_2(t) = 0$, then condition~(\ref{eq01}) is directly satisfied.
Inserting~(\ref{defphi2}) in Eq.~(\ref{eq02}) and integrating by part, we get for $r>r_c$
\begin{align*}
f_2(r)=-  \Phi_2(r_c) & \int_0^{\infty} \d q \,   J_0(qr)  \sin (qr_c)   \\
& - \int_{r_c}^{\infty} \d t \,   \Phi_2'(t) \int_0^{\infty} \d q \,  J_0(qr) \sin (qt)   \ .
\end{align*}
Then using~(\ref{intsin}), we  obtain the integral equation
\begin{equation*}
f_2(r)=- \int_{r}^{\infty}  \frac{\Phi_2'(t)}{\sqrt{t^2-r^2}}    \d t   \ ,
\end{equation*}
so that we finally get for all $t \geq r_c$
\begin{equation}
\Phi_2(t) = \frac{2}{\pi}  \int_{t}^{\infty}  \frac{rf_2(r)}{\sqrt{r^2-t^2}}   \d r   \ .
\label{resphi2}
\end{equation}

\subsection{General solution}

The conclusion is now straightforward: once the functions $f_1$ and $f_2$ are specified, both auxiliary functions $\Phi_1$ and $\Phi_2$ can be evaluated according to~(\ref{resphi1}) and~(\ref{resphi2}). Finally, the total amplitude $A(q)  = A_1(q) + A_2(q)$ is  obtained by integrating~(\ref{defphi1}) and~(\ref{defphi2}).

\section{Numerical simulations}
\label{appB}

The simulations are performed in two-dimensional axisymmetrical geometry. 
We first write the transport equations in non-dimensional form. Define $a$, $V_{\textrm{inj}}$, $\eta V_{\textrm{inj}}/a$, and $\Gamma_0$ respectively as length, velocity, pressure, and concentration scales, one arrives at
\begin{align}
& \nabla^2 \mathbf{v} = \bm{\nabla} p \ , \quad \bm{\nabla} \cdot \mathbf{v} = 0 \ , \label{b1} \\
& \text{Pe}_{\textrm{num}} \bm{\nabla} \cdot (\mathbf{v} \Gamma) = \nabla^2 \Gamma  \ , \label{b2}
\end{align}
where the P\'eclet number is defined as $\text{Pe}_{\textrm{num}}=aV_{\mathrm{inj}}/D$. The velocity and concentration fields are coupled through the Marangoni boundary condition. In dimensionless form, the latter becomes
\begin{equation}
\beta_{\textrm{num}}^{\infty} \partial_z v_r \big\vert_{z=0} = - \frac{x}{1-x \Gamma} \partial_r \Gamma \ ,
\label{b3}
\end{equation}
with $x=\Gamma_0/\Gamma_{\infty}$, and $\beta_{\textrm{num}}^{\infty} = \eta V_{\textrm{inj}}/(\Gamma_{\infty} k_B T)$.  In this work, we arbitrarily set $\text{Pe}_{\textrm{num}}=3.57\times 10^3$ and $\beta_{\textrm{num}}^{\infty} = 1.06 \times10^{-3}$.

The total typical number of elements for the small (resp. large) cell was around 25000 (resp. 45000). We check that increasing the number of elements had insignificant quantitative consequences for the computed quantities of interest.
We use the Laminar Flow module combined with the Coefficient Form Boundary PDE module of COMSOL to solve for the fluid flow transport equations in the bulk [Eq.~(\ref{b1})] and the transport of insoluble surfactants at the free surface [Eq.~(\ref{b2})] together with the associated boundary conditions [in particular Eq.~(\ref{b3})]. Since the geometry is 2D axisymmetric, care is taken to compensate for the missing terms between the covariant differentiation of the divergence and laplacian operators in Eq.~(\ref{b2}) and the regular partial differentiation that the COMSOL PDE module considers by default (see e.g.  \textsf{https://www.comsol.com/blogs/guidelines-for-equation-based-modeling-in-axisymmetric-components/}). 

We discretize the fluid flow with quadratic elements for the velocity field and linear elements for the pressure field; quadratic elements are employed to discretize the interfacial concentration field. We use either the MUMPS or PARDISO solver to obtain the steady-state of the system, which is typically reached   after $\sim 10\,$mn (physical time).

\section{Relation between $V_0$ and $V_{\textrm{inj}}$}
\label{appC}

\begin{figure}[b]
\centering
\includegraphics[width=0.8\columnwidth]{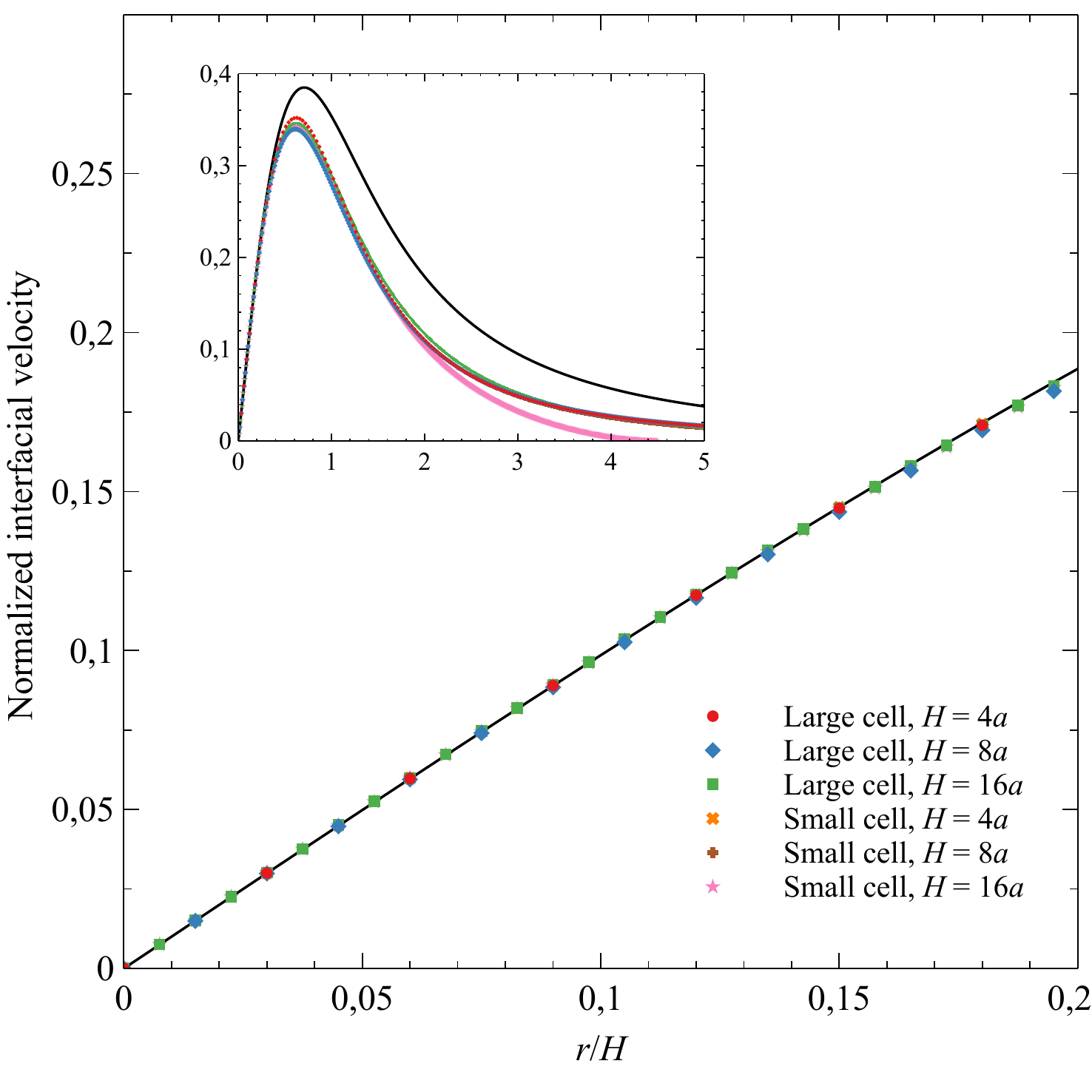}
\caption{Normalized interfacial velocity $v_r(r,0)/[V_{\textrm{inj}} \times h(H/a)]$ as a function of the distance to the origin. The full line corresponds to  Eq.~\ref{c2}. Inset: same data presented over a wider range.}
\label{fig10}
\end{figure}

When comparing the numerical data  with the analytical predictions, we are facing the difficulty that the velocity scales are not defined in the same manner. The flow is due to a point source of momentum in the theory, whereas the jet velocity is prescribed in the simulations. It is therefore legitimate to wonder what is the relation between the quantity $V_0$ introduced in Eq.~(\ref{v0})  and the injection speed $V_{\textrm{inj}}$. To answer this question, we consider the interfacial flow in the case of a pure interface ($\Gamma_0=0$). Since this study is restricted to the Stokes regime ($\text{Re}=0$), one expects a universal relation of the form
\begin{equation}
v_r(r,0) = V_{\textrm{inj}} \times f \left(r/H,a/H \right)  \ ,
\label{c1}
\end{equation}
with $a$ the radius of the injection nozzle. One the other hand, the theory Eq.~(\ref{v0}) predicts 
\begin{equation}
 v_r (r,0) =\frac{V_0}{2} \frac{r/H}{\left[ \left(r/H\right)^2 +1 \right]^{3/2}}    \ .
 \label{c2}
\end{equation}
One thus expects $f$ to scale as $f \left(r/H,a/H \right) \sim r/H$ when $r \to 0$ (remember that $a=0$ in the analytical description). We then define a new  function $g$ such that $f(x,y)=xg(x,y)$. Comparing Eqs.~(\ref{c1}) and~(\ref{c2}), one gets
\begin{equation}
 \frac{V_0}{2 V_{\textrm{inj}}} = g(0,a/H) \doteq  h(H/a)    \ .
 \label{c3}
\end{equation}
The issue is then to characterize the universal function $h$ from the numerical data. To this aim, we investigate two cell sizes ($R=72a$ and $R=144a$) and three values for the gap ($H=4a$, $8a$ and $16a$). The behavior of $h(H/a)$ is figured out by extracting the limit of $Hv_r(r,0)/(rV_{\textrm{inj}})$ when $r \to 0$ (\textit{i.e.}, we evaluate the slope of the velocity profile at the origin). It can be checked that this limit is independent of the cell size (see Fig~\ref{fig10}), and we obtain: $h(4) \approx 0.237$, $h(8) \approx 0.091$, and $h(16) \approx 0.035$. Interestingly,  the ratio of consecutive values is (almost) constant: $h(4)/h(8)\approx h(8)/h(16) \approx 2.60$. This leads us to suggest the following power law
\begin{equation}
h(x)  = K x^{-\alpha}   \ ,
 \label{c4}
\end{equation}
with $K$ and $\alpha$ two fitting parameters whose numerical values are: $\alpha \approx 1.38$ and $K \approx 1.60$.

This discussion advocates that, even though $V_0$ and $V_{\textrm{inj}}$ must be proportional to each other (as it should be at low Reynolds number), their ratio actually depends non-trivially on the gap~$H$. As shown in Fig.~\ref{fig10}, the fitting procedure described above works extremely well in the vicinity of the origin. At larger distances, however, the agreement with the expression~(\ref{c2}) is not as accurate, presumably due to finite-size effects (either because the cell size is not infinite, or because Eq.~(\ref{c2}) refers to the limit $a\to 0$ in the function $f(r/H,a/H)$).


\begin{thebibliography}{99}
\bibitem{remark} 
Strictly speaking, the asymptotic value of the concentration should be defined as $\Gamma_0 + \delta \Gamma_0$, where the excess  value $\delta \Gamma_0$ is set by mass conservation as well. It can actually be shown that $\delta \Gamma_0 \sim 1/R$, with $R$ the size of the system. The correction is thus irrelevant in the limit $R \to \infty$.
\end{thebibliography}
\end{document}